\documentclass[%
 reprint,
 twocolumn,
 floatfix,
 superscriptaddress,
 amsmath,amssymb,
 aps,prx
]{revtex4}

\usepackage{graphicx}
\usepackage{dcolumn}
\usepackage{bm}

\usepackage[hyperindex,breaklinks]{hyperref}
\usepackage{amsmath}
\usepackage{graphicx}
\usepackage{CJK,braket}
\usepackage{mathrsfs}
\usepackage{caption}
\usepackage{subfig}

\newcommand{\B}{\mathcal{B}}

\def\be{\begin{equation}}
\def\ee{\end{equation}}
\def\ba{\begin{eqnarray}}
\def\ea{\end{eqnarray}}

\begin{document}


\title{Genuine Quantum Chaos and Physical Distance Between Quantum States}
\begin{CJK}{UTF8}{gbsn}
\author{Zhenduo Wang (王朕铎)}
\affiliation{International Center for Quantum Materials, School of Physics,
Peking University, 100871, Beijing, China}
\author{Yijie Wang (王一杰)}
\affiliation{International Center for Quantum Materials, School of Physics,
Peking University, 100871, Beijing, China}
\author{Biao Wu(吴飙)}
\affiliation{International Center for Quantum Materials, School of Physics,
Peking University, 100871, Beijing, China}
\affiliation{Wilczek Quantum Center, School of Physics and Astronomy,
Shanghai Jiao Tong University, Shanghai 200240, China}
\affiliation{Collaborative Innovation Center of Quantum Matter, Beijing 100871,  China}
\date{\today}


\date{\today}


\begin{abstract}
We show that there is genuine quantum chaos despite that quantum dynamics is linear. This is revealed by introducing
a physical distance between two quantum states. Qualitatively different from existing distances for quantum states, for example,
the Fubini-Study distance, the physical distance between two mutually orthogonal quantum states can be very small.  As a result,
two quantum states, which are initially very close by physical distance, can diverge from each other during the ensuing quantum
dynamical evolution. We are able to use physical distance to define quantum Lyaponov exponent and quantum chaos measure.
The latter  leads to quantum analogue of the classical Poincar\'e section, which maps out the regions where quantum dynamics
is regular and the regions where quantum dynamics is chaotic.
Three different systems, kicked rotor, three-site Bose-Hubbard model, and spin-1/2 XXZ model, are used to illustrate our results.
\end{abstract}


\maketitle
\end{CJK}

\section{\label{sec:1}Introduction}
\begin{figure}[tbp]
\centering
\includegraphics[width=0.85\columnwidth]{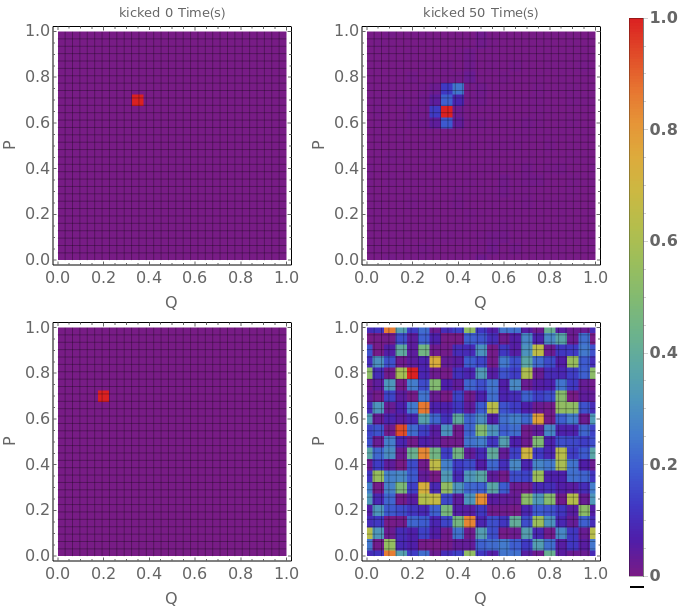}
\caption{Two distinct dynamical evolutions of quantum kicked rotor with kicking strength $K=4.7$ and resolution $m=20$, i.e., effective Planck constant $\hbar_{\text{eff}} = 2\pi/m^2\approx 0.016$. These two dynamical evolutions are generated by the same Hamiltonian but with different initial conditions(different Wannier basis states). See \hyperref[subsec:5-1]{V. A} for details  }
\label{kicked}
\end{figure}

As classical equations of motion are in general nonlinear, there are mainly two types of classical motion,
regular motion that does not depend sensitively on initial conditions and chaotic motion that does~\cite{arnol2013mathematical}.
In contrast,  the Schr\"odinger equation is linear,  and it is widely believed that there is no true chaotic motion in
quantum dynamics ~\cite{Berry1989,Weinberg}. However, this belief
contradicts  the fact that there are also two types of quantum motion, regular and chaotic. Shown in Fig. \ref{kicked}
is one example.  With two different initial conditions that are both well localized, quantum kicked rotor exhibits two very different dynamics:
after 50 kicks, one wave packet remains well localized; the other  spreads out
widely with an irregular pattern.

\begin{figure}[tbp]
\centering
\includegraphics[width=0.85\columnwidth]{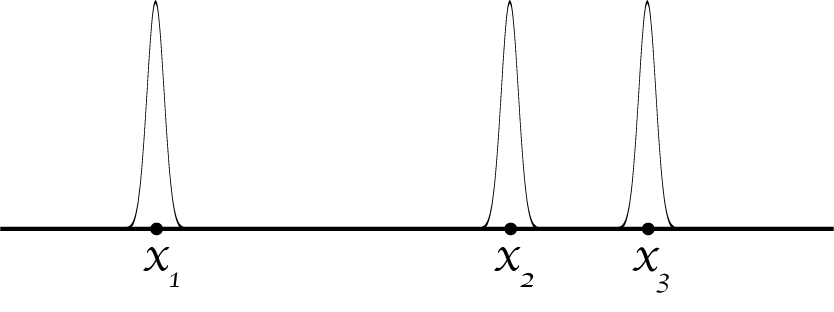}
\caption{Three well localized wave packets  at positions $x_1$, $x_2$ and $x_3$, respectively.
 There is no overlap between these wave packets. Physically, the two wave packets on the right are closer to each other. }
\label{g3}
\end{figure}

This widely-held misunderstanding is rooted in that people use the inner product $\braket{\psi_1|\psi_2}$ to measure
the difference between two quantum states $\ket{\psi_1}$ and $\ket{\psi_2}$, such as in Fubini-Study distance\cite{Fubini}
and many others \cite{Adjisavvas1981,Hillery1987,Luo2004,Filippov2010,Rana2016,Wootters1981,Braunstein1994}.
As a result, two mutually orthogonal quantum states
always have the same distance. This is clearly inadequate in at least two aspects. ({\it i}) These inner-product based distances
do not reduce to the distance between two classical states at the semiclassical limit. ({\it ii}) These
distances are not consistent with our physical intuition in many familiar situations.
One example is shown in Fig.\ref{g3}, where there are three well localized wave packets that are orthogonal to each other.
It is intuitively evident that the the wave packet at $x_2$ is  physically closer to the one at $x_3$ than the one at $x_1$.
Another example is a one dimensional spin chain. Suppose that we have  three states
$\ket{\phi_1}=\ket{1,1,1,\cdots,1}$, $\ket{\phi_2}=\ket{-1,1,\cdots,1}$, and $\ket{\phi_3}=\ket{-1,-1,\cdots,-1,1,1,\cdots,1}$, which are
orthogonal to  each other. It is clear that $\ket{\phi_1}$ and $\ket{\phi_2}$ are very
closely to each other physically as they have almost the same
magnetization while $\ket{\phi_1}$ and $\ket{\phi_3}$ are very different to each other physically.
Hamming distance would be more appropriate. Recently, some other works have also tried to go
beyond the inner product  to quantify the difference between quantum states\cite{yan2020quantum,Chen2018}.

In this work we show that one can distinguish the two different types of quantum motions shown in Fig.\ref{kicked} by
introducing a physical  distance between quantum states based on the Wasserstein distance.
Due to the use of  the distance defined between basis vectors, our quantum distance is capable of quantifying the
physical difference between quantum states. In particular, (1) it can reduce to the distance between classical states
at the semiclassical limit; (2) it is not conserved during the quantum dynamical evolution; (3) it 
can be  small or large between a pair of mutually orthogonal quantum states. This is qualitatively different from existing distances defined
between quantum states, for example, Fubini-Study distance\cite{Fubini}.
As a result, two quantum states, which are orthogonal to each other and initially close in the physical distance,
can dynamically diverge from each other in the physical distance
despite that the inner product stays at zero (see more detailed discussion at the beginning of Section \ref{sec:4}). 
This physical distance allows us to define two parameters, quantum Lyapunov exponent and quantum chaos measure,
to characterize quantum motion.  Specially,  the quantum chaos measure can be used to construct the quantum analogue of the classical Poicar\'e section,
where we can map out the regions, where the quantum motion is regular (e.g., see Fig.\ref{fig2}),
and the regions, where the quantum motion is chaotic and depends sensitively on the initial condition (e.g., see Fig.\ref{fig2}).
This quantum Poicar\'e section reduces to its classical counterpart at the semiclassical limit.

We will introduce our  definition of quantum physical distance in Sec. \ref{sec:2}.
The soundness and usefulness of our distance  is then illustrated with  examples in Sec. \ref{sec:3}.
In Sec.  \ref{sec:4}, with quantum physical distance, we define two parameters, quantum Lyapunov exponent
and quantum chaos measure. The former  characterizes the short-time dynamical behavior of a quantum state while
the latter the long-time dynamical behavior of a quantum state. These concepts are numerically illustrated with
three different quantum systems in Sec. \ref{sec:5}, which include the kicked rotor as the system which
has a clear classical counterpart, a three-site Bose-Hubbard model whose classical counterpart is a
mean field theory, and the spin-chain which does not have an obvious classical counterpart.
Finally we discuss and conclude.

\section{\label{sec:2}Physical Distance Between Quantum States}

Our physical distance between quantum states is based on the Wasserstein distance, which is a distance function defined
between probability distributions on a metric space.
In computer science it is known as the earth mover's distance and has been widely used in many fields\cite{Rubner2000,10.1371/journal.pone.0151859}.
To define a Wasserstein distance,  we need both a metric space and a distribution function. To have them, for a quantum system,
we choose a complete set of orthonormal basis $\B=\{\ket{\xi_1},\cdots,\ket{\xi_n}\}$ and define a distance between
the bases $d(\xi_i,\xi_j)$. This gives us a metric space.
When a given quantum state $\ket{\psi}$ is expanded in terms of this basis, we have a probability distribution on the set \(\B\)
\begin{equation}
  p_i(\psi) = |\bra{\xi_i}\psi\rangle|^2 \,,~~~~ \ i=1,\cdots,n \label{eq:1}\,.
\end{equation}
Our physical distance between two quantum states \(\ket{\psi_1},\ket{\psi_2}\) is the Wasserstein-\(\lambda\) distance between
distributions $p_i(\psi_1)$ and $p_j(\psi_2)$
\begin{equation}
\label{phyd}
    D_\lambda(\psi_1,\psi_2) = \Big[\inf_{P} \sum_{i,j} P_{ij} d^\lambda(\xi_i,\xi_j)\Big]^{1/\lambda}
\end{equation}
where  $\lambda$ is a positive integer and $\inf_{P}$ means the minimum over all the distributions $P_{ij}\in [0,1]$ that satisfy
\be
\sum_{i=1}^n P_{ij} = p_j(\psi_2) \ ; ~~~~\sum_{j=1}^n P_{ij} = p_i(\psi_1)\,.
\ee
It is clear that the above definition still works even when $n$ is infinite. For most of the cases studied in this work, we choose $\lambda=1$.
This definition of physical distance can be generalized straightforwardly for mixed states. To do it, one only needs to
specify the probability distribution as $p_i(\hat \rho) = \text{Tr} (\ket{\xi_i}\bra{\xi_i}\hat \rho)$ for a mixed state
described by density matrix $\hat \rho$.

Two points warrant attention. (1) For a given quantum system, the choice of the orthonormal basis $\B$ is not unique.
It depends on the physical issue that people want to address. For example, for a spin-lattice system, if we are interested in the magnetization along a given direction,
then the spin up and down states in that direction are a natural choice and the distance $d$ for the metric can be chosen as the Hamming distance. (2)
Our physical distance \(D_\lambda\) is not a distance on the Hilbert space \(\mathcal{H}\).
 There exists the states $\ket{\psi_1}\neq \ket{\psi_2}$ for whom $D_\lambda(\psi_1,\psi_2)=0$, for example,  $\ket{\psi_1} = (\ket{\xi_1}+\ket{\xi_2})/\sqrt{2}$ and $\ket{\psi_2} = (\ket{\xi_1}-\ket{\xi_2})/\sqrt{2}$.  Therefore, our distance is a function of states and basis, i.e., quantum states and the way
 to extract physical information from them. More thorough discussion will be given with examples in the following sections.

In Ref. \cite{Zyczkowski1998,Zyczkowski_2001}, a  Monge distance was defined between quantum states with
 the Husimi function of a quantum state as the distribution. It shares two features with our distance: (1) mathematically,
 both are Wasserstein distance; (2) both reduce to the distance between classical states in the semiclassical limit $\hbar\rightarrow 0$.
 However, there is a crucial difference: the use of the orthonormal basis $\B$ and a metric defined over $\B$ in our definition.
 As a result, our  physical distance is applicable for all quantum systems, including spin systems.
If one is forced to view the Monge distance in this perspective,  its choice of   the orthonormal basis $\B$ is the points in 
the classical phase space and 
 the metric is the usual distance between these points. This choice is certainly not natural as the Monge distance
 is defined for quantum states.

\section{\label{sec:3}Examples of Physical Distance}
In this section, we use a few examples to illustrate the physical distance between quantum states. We will see that
it can indeed capture quantitatively the physical difference between quantum states and is consistent with our physical intuition.
There are various distances between quantum states based on the inner product of quantum states; for the sake of convenience,
we compare our physical distance to one of them, Fubini-Study distance~\cite{Fubini}.

The first example is a one-dimensional spinless particle and we are interested in its position. In this case, the basis $\B$ consists of infinite number
of vectors $\ket{x}$, which are eigen-functions of position operator $\hat{x}$. We define the  distance $d$  between two basis vectors $\ket{x}$
and $\ket{x^\prime}$ as $d(x,x^\prime)=|x-x^\prime|$. Consider two different quantum states, $\ket{x_1}$ and $\ket{x_2}$.
Then according to our definition, the physical distance between them is $D_2(x_1,x_2)=|x_1-x_2|$.
In contrast, the Fubini-Study distance between $\ket{x_1}$
and $\ket{x_2}$ is one as long as $x_1\neq x_2$. Let us consider a Gaussian wave packet,
\begin{equation}
  \braket{x|\psi_{x_0,p_0;\sigma}}=\frac{1}{(2\pi\sigma^2)^{\frac14}} \exp\Big[-\frac {(x-x_0)^2} {4\sigma^2} + i \frac {xp_0} {\hbar}\Big]\,.
\end{equation}
One can find that the physical distance between two different Gaussian states is~\cite{Olkin1982}
\begin{equation}
  D_2(\psi_{x_1,p_1;\sigma_1},\psi_{x_2,p_2;\sigma_2}) = \sqrt{(x_1-x_2)^2+(\sigma_1-\sigma_2)^2}\,.
\end{equation}
If the two Gaussian wave packets have the same width $\sigma_1=\sigma_2$, we simply have $D_2= |x_1-x_2|$,  which is just what our
physical intuition expects. In contrast, the Fubini-Study distance between these
two Gaussian packets is close to one as long as $ |x_1-x_2|\gg \sigma_{1,2}$. It is interesting to note that $D_2$ is independent of the momentum.
This is reasonable as we are currently interested in the particle's position. If one is interested in the particle's momentum,
one can similarly define $d(p,p^\prime)=|p-p^\prime|$ and then find the physical distance in momentum between two Gaussian packets
as
\begin{equation}
\label{qpdis}
  \widetilde{D}_2(\psi_{x_1,p_1;\sigma_1},\psi_{x_2,p_2;\sigma_2}) = \sqrt{(p_1-p_2)^2+(\tilde{\sigma}_1-\tilde{\sigma}_2)^2}\,,
\end{equation}
where $\tilde{\sigma}_{1,2}$ are the widths of the wave packets in the momentum space.

\begin{figure}[htbp]
\centering
\includegraphics[width=0.55\columnwidth]{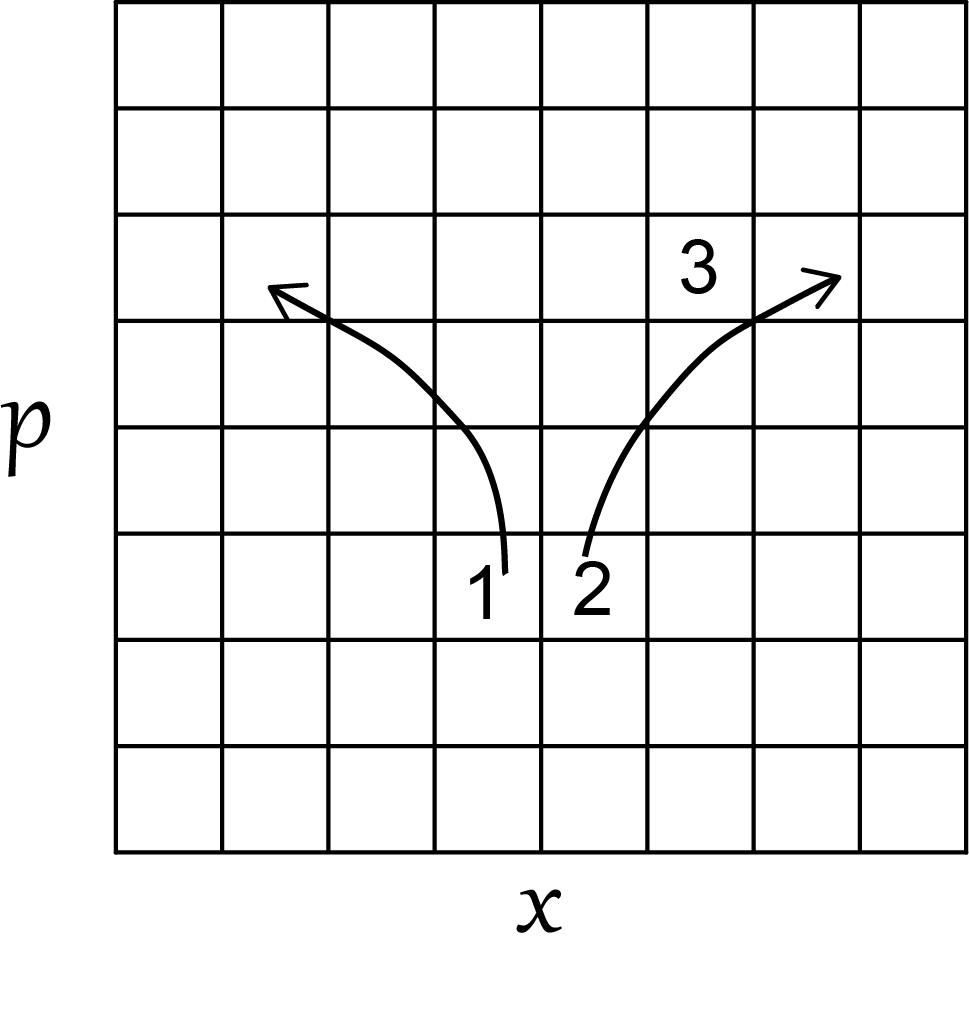}
\caption{Quantum phase space of a one dimensional particle. $p$ and $x$ are its momentum and position, respectively.
Each square represents a Planck cell. Three different Planck cells are marked by 1, 2, and 3. }
\label{phase}
\end{figure}

The above simple example shows that the physical distance depends on what physics we want to explore. Mathematically, this is achieved by
choosing an appropriate set of orthonormal basis $\B$.  If we want to explore physics that is related explicitly to
both position and momentum, we can choose $\B$ to be a set of Wannier basis. As shown in Fig.\ref{phase}, the classical phase space
is divided into Planck cells and each Planck cell is
assigned a Wanner function $\ket{w_{j}}$~\cite{Neumann1929,vonNeumann2010qhtheorem,Han2015,Fang}.
These orthonormal Wanner functions $\ket{w_{j_xj_p}}$ form the basis $\B$. We define
\begin{equation}
d(w_{j_1},w_{j_2})=\sqrt{(x_{j_1}-x_{j_2})^2+(p_{j_1}-p_{j_2})^2}\,,
\end{equation}
where $x_j$'s and $p_j$'s are the coordinates of the Planck cells $\ket{w_j}$'s.  Let us consider two Gaussian packets
of the same width $\ket{\varphi_1}$ and $\ket{\varphi_2}$ in the quantum phase space, $\ket{\varphi_1}$ is
centered at $\ket{w_1}$ and $\ket{\varphi_2}$ centered at $\ket{w_2}$.  When the widths of two packets
are  much larger than a Planck cell and much smaller than the  distance $d(w_{j_1},w_{j_2})$,  we should have
\begin{equation}
  \widetilde{D}_1(\phi_1,\phi_2) \approx \sqrt{(x_1-x_2)^2+(p_1-p_2)^2}\,,
\end{equation}
where the approximation is due to that the Gaussian packets are  discretized in the quantum phase space.
This physical distance is reduced to the distance in the classical phase space when $\hbar\rightarrow 0$.
The Fubini-Study distance\cite{Fubini} does not have this kind of semi-classical limit. 

We turn to many-body quantum states, and  choose Fock states as the basis $\B$. A quantum state
$\ket{n_j}=\ket{n_1,n_2,\cdots,n_k}$ means that there are $n_j$ particles in the  single particle mode $\ket{e_j}$.
The vacuum state is specially denoted as $\ket{e_0}=\ket{0,0,\cdots,0}$.  We define a metric for
the single-particle modes and the vacuum mode as $d(e_i,e_j) = d_{ij}=d_{ji}$ and $d(e_i,e_0) = d_{i0}= d_{0i}$.
This allows us to define the distance between two Fock states $\ket{n_j}$ and $\ket{m_j}$
\begin{equation}
d(\ket{n_j},\ket{m_j}) = \text{min}_{\Delta_{ij}} \sum_{i,j} d_{ij}\Delta_{ij}
\end{equation}
where $n_i = \sum_j \Delta_{ij}$, $m_j = \sum_i \Delta_{ij}$, and $\Delta_{ij} \geq 0$. If the total number of particles in these two Fock states
are different, we let  $m_0 = \max(\sum_i n_i, \sum_i m_i) - \sum_i m_i$ or $n_0 = \max(\sum_i n_i, \sum_i m_i) - \sum_i n_i$
be the occupation number for the vacuum state $\ket{e_0}$. As a result,
our definition is legal for states of different particle numbers.  Note that the distance $d_{ij}$ can be defined differently for different systems
and different physics that one is interested.  Our definition  has at least two advantages. First, it shares the same
spirit with our physical distance, it is a Wasserstein-like metric for particle number distributions. Second, there is no exponential scaling between distance and particle number,
which exists in the Fubini-Study distance.

We use a special case to illustrate the second point.
Consider a  system of $N$ identical Bosons and its two quantum states. In one state $\ket{\Psi_1}$, all the Bosons are in the mode
$\ket{e_1}$; in the other state $\ket{\Psi_2}$,  all the Bosons are in the state $\alpha\ket{e_1}+\beta\ket{e_2}$.
It can be shown that $D_1(\Psi_1,\Psi_2)\propto N$. In contrast, the Fubini-Study distance is about $1-|\alpha|^{2N}$, which can be regarded
as one when $N$ is large even when $\alpha\sim 1$, reflecting the fact that the two many-body states $\ket{\Psi_1}$ and $\ket{\Psi_2}$
are almost orthogonal to each other when $N$ is large no matter how close the single particle states $\ket{e_1}$
and $\alpha\ket{e_1}+\beta\ket{e_2}$ are  to each other. So, our physical distance is more consistent with our  intuition.

\section{\label{sec:4}Quantum Dynamics and Physical distance}
 As quantum dynamics is linear,  it is often said that there is no true  chaos, in the sense of the chaos seen in nonlinear classical dynamics~\cite{Berry1989,Weinberg}.
The argument is as follows.
Suppose that we have two quantum states $\ket{\psi_1}$ and $\ket{\psi_2}$. As quantum dynamics is linear, the inner product
$\braket{\psi_1|\psi_2}$ does not change with time.  As a result, if these two states $\ket{\psi_1}$ and $\ket{\psi_2}$ are very close to each other,
that is, $\braket{\psi_1|\psi_2}\sim 1$, they will always be close to each other.  This implies  no true chaos. However, as shown in Fig.\ref{kicked},
chaotic motion is clearly possible in quantum dynamics. Other examples of chaotic quantum motion can be found in Ref.~\cite{Han2016}.
 This contradiction is due to the use of the inner product to measure the
difference between quantum states. As already discussed, the inner product is incapable of telling us how close or far two quantum states
are when they are orthogonal to each other.

Let us consider the case in Fig. \ref{phase}. We use $\ket{w_1}$ and $\ket{w_2}$ to denote the two quantum states represented by
the two Planck cells marked with 1 and 2. We let $\ket{w_1}$ and $\ket{w_2}$ to evolve, respectively,  according to a given Schr\"odinger
equation. As a result, at time $t$,  $\ket{w_1}$ becomes $\ket{\phi_1(t)}$ and $\ket{w_2}$ evolves into $\ket{\phi_2(t)}$.
The Fubini-Study distance between these two states does not change with time as $\braket{\phi_1|\phi_2}=\braket{w_1|w_2}=0$. However,
it is very different story for physical distance. According to our discussion, the physical distance between $\ket{w_1}$ and $\ket{w_2}$ is small.
When the dynamical evolution starts, the physical distance can grow. The linearity of quantum dynamics does not guarantee   that
the physical distance between $\ket{\phi_1(t)}$ and $\ket{\phi_2(t)}$ be small. The situation shown in Fig. \ref{phase} can happen:
the physical distance grows with time in quantum dynamics while keeping $\braket{\phi_1|\phi_2}$ at zero. As we will show with our
numerical calculation in the next section, this is indeed what happens in quantum chaotic systems.

So,  there is true chaos in quantum dynamics and it can be understood in terms of physical distance.
In the following, using the concept of physical distance,
we define two parameters to characterize the diverging and irregular quantum dynamics.

\subsection{\label{subsec:4-1}Quantum Lyapunov Exponent}

Lyapunov exponent is one of the most important concepts in classical dynamics and it characterizes the rate of separation of infinitesimally close
initial trajectories. Quantitatively, in a chaotic classical dynamics,
the distance (usually  $L_2$ distance) between two points that are initially very close grows with time $t$ as
\begin{equation}
  \|\delta \bm{Z}(t)\| \approx e^{\gamma t} \|\delta \bm{Z}(t=0)\|
\end{equation}
where $\bm{Z} = (q,p)$ is the state in phase space. The parameter $\gamma$ is the Lyapunov exponent.
With the physical distance between two quantum states, the (maximum) Lyapunov exponent in quantum mechanics can be similarly defined as\cite{MANKO2000330,ZYCZKOWSKI1993153}
\begin{equation}
  \gamma_Q = \lim_{t\rightarrow \infty} \lim_{\psi' \rightarrow \psi} \frac 1 t \log\frac {D(\psi(t), \psi'(t))} {D(\psi(0), \psi'(0))}
\end{equation}
This is very similar to the definition of Lyapunov component in classical mechanics, which can be obtained by replacing
the physical distance between states $D(\cdot, \cdot)$ with the distance in classical phase space.
The symbol $\psi'\rightarrow \psi$ means $\psi,\psi'$ are close but different in the sense of physical distance.

For  quantum systems where quantum phase spaces similar to Fig.\ref{phase} can be constructed,
we can always use physical distance similar to the one in Eq.(\ref{qpdis}).  In the semi-classical limit, $\hbar\rightarrow 0$, the areas or volumes of
the Planck cells approach zero and the quantum dynamics becomes classical. In this limit, we should have
\begin{equation}
  \lim_{\hbar \rightarrow 0} \gamma_Q = \gamma_C\,.
\end{equation}
Note that this relation holds only when the Lyapunov  time (the inverse of Lyapunov exponent) is smaller than the Ehrenfest time~\cite{Han2016,Zhao2019}.
So the limit $t\rightarrow \infty$ is not a strict mathematical term and should be understood as  a sufficiently long time
before the wave packets become too  widely spread.

\subsection{\label{subsec:4-2}Quantum Chaos Measure}

Intuitively,  chaos means disorder and irregularity in dynamics.
In classical dynamics, this is indicated by the scattered points in Poincar\'e sections (see, e.g., Fig.\ref{fig2}(a)), which are usually
referred to as chaotic sea. How  chaotic a classical dynamics is reflected by how much the chaotic seas occupy in the phase space.
In the chaotic sea, there are regular motions, which are usually referred to as integrable island.
When there is only ``chaotic sea" in the Poincar\'e sections, the system
becomes fully chaotic   In this case, we have ergodicity and/or mixing and the long-time average becomes identical to the microcanonical
ensemble average~\cite{qmixing}.

With physical distance, we can also compare the long-time average and the microcanonical
ensemble average for quantum dynamics. In standard textbooks~\cite{Huang},  the microcanonical
ensemble is regarded as a maximally mixed state and can be described by the density matrix $\hat{\mathbb{I}}/{\rm Dim}$.
$\hat{\mathbb{I}}$ is the identity matrix and Dim is the dimension of the Hilbert space.
This is usually a postulate in standard textbooks~\cite{Huang}. But it has been fully justified by many
studies~\cite{Neumann1929,vonNeumann2010qhtheorem,reimann2008,Han2015,qmixing}.
We use the physical distance between the long-time-average of the density matrix and the density matrix $\hat{\mathbb{I}}/{\rm Dim}$ to
quantitatively measure the severity of quantum chaos. Mathematically, this difference is given by
\begin{equation}
  \Upsilon = D\Big(\frac 1 {\text{Dim}}, \lim_{T\rightarrow\infty}\frac 1 T\int_0^T \hat \rho(t)\text{d} t  \Big)
\end{equation}
We call it quantum chaos measure. The measure $\Upsilon$ depends on the initial quantum states. For some initial quantum states,
 $\Upsilon$ is small and it means that the long-time-average density matrix is sufficiently close to the maximally mixed state. These quantum states
 belong to chaotic sea. For some initial states, $\Upsilon$ is large and these states belong to integrable islands. In the next section, our numerical results
 will show that we can use chaos measure to construct quantum Poincar\'e sections, which resemble classical Poincar\'e sections.

One can use other tools to quantify the degree of disorder in quantum dynamics, for example, quantum entropy of the
form $-\sum_i p_i \log p_i$~\cite{Jiang2017,Hu2019}.  The core advantage of our chaos measure is its dependence on the metric structure of basis $\B$, i.e., the information of base space.
For example (see Appendix \ref{app-2} for detail), consider the following two probability distributions on set $\{0, 1, 2, \cdots, 9\}$ 
\begin{equation}
  p_A(x) = \begin{cases}
  1/5 & x < 5\\
  0 & x \geq 5
\end{cases}
\end{equation}
and
 \begin{equation}
  p_B(x) = \begin{cases}
  1/5 & x \text{ is even}\\
  0 & \text{otherwise}
\end{cases}
\end{equation}
The entropies of $p_A$ and $p_B$ are the same, but $p_B$ appears much closer to the uniform distribution.  This can be reflected by our chaos measure
as we have $\Upsilon_A = 5/2 \ ; \ \Upsilon_B = 1/2$ with the metric $d(x,y) = |x-y|$ on the base space. This difference means our measure $\Upsilon$ can
reveal finer property better than any other concepts that ignore the information of the base space. In Ref.\cite{Yin2019},  the length of a Planck cell 
was introduced to measure disorder
in quantum dynamics; however, this concept is limited and can not be applied to spin systems.




\section{\label{sec:5}Numerical Results}
In this section, we will numerically study three different systems  to illustrate the concept of physical distance.
These three systems are quantum kicked rotor, three-site Bose-Hubbard model, and  XXZ spin chain. The quantum kicked rotor
has a natural classical counterpart. For the three-site Bose-Hubbard model, its classical counterpart is the mean-field theory and
its effective Planck constant is the inverse of the particle number
$1/N$.  In contrast, the   XXZ spin chain has no obvious classical counterpart.

\subsection{\label{subsec:5-1}Kicked Rotor}
Kicked rotor is one of the systems which have been well studied both as a quantum and classical system.
The Hamiltonian of a kicked rotor on  a ring has the following dimensionless form~\cite{Jiang2017,Yin2019}
\begin{equation}
  H = \frac 1 2 p^2 + K \cos q \sum_{n=-\infty}^{+\infty} \delta(t-n)
\end{equation}
Its classical dynamics is equivalent to the following map
\begin{equation}
  \begin{aligned}
    p_{n+1} &= p_n+K \sin q_n \mod 2\pi \\
    q_{n+1} &= q_n+p_{n+1} \mod 2\pi
  \end{aligned}
\end{equation}
where we have used the fact that the momentum $p$ and $p+2n\pi$ are equivalent. $(q_n,p_n)$ is the
position and momentum of the kicked rotor before the $n$-th kick.  The kicking strength $K$ is
the only control parameter; when it is bigger than $K_c=0.971635$ the classical dynamics
becomes chaotic\cite{ChirikovScholarpedia}.

The quantum dynamics has one more parameter, the effective Planck constant $\hbar_{\rm eff}$~\cite{Jiang2017,Yin2019}.
For simplicity, we choose $\hbar_{\rm eff}=2\pi / m^2$ with $m$ being an positive integer. In this case, we can divide
the $2\pi\times 2\pi$ classical phase space into $m \times m$ Planck cells (similar to Fig. \ref{phase}) and assign
a Wannier function $\ket{X,P}$ to each Planck cell~\cite{Jiang2017,Yin2019}.  $X,P$ are the coordinates of a Planck cell.
These Wannier functions $\{\ket{X,P}\}$ form a complete set of orthonormal basis.  We choose them as  our choice of $\B$
and define the distance between two basis vectors as $d(X,P;X',P') = \sqrt{(X'-X \mod 2\pi)^2+(P'-P \mod 2\pi)^2}$.



\begin{figure}[htbp]
  \includegraphics[width=\columnwidth]{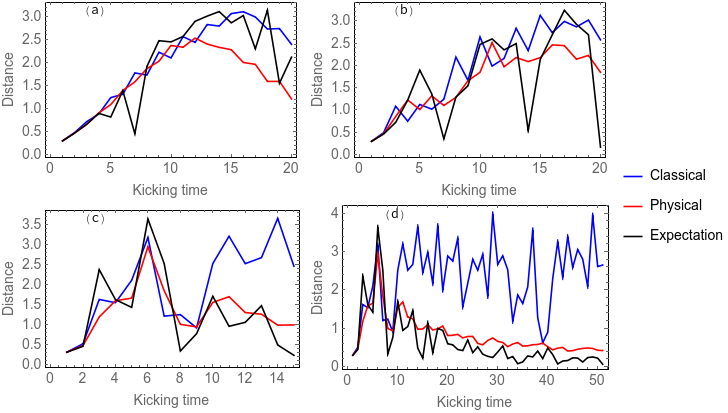}
  \caption{\label{fig1}The time evolution of three types of distances for the kicked rotor at different kick strengths $K$. Blue lines are the distances between the points on classical trajectories, red lines are physical distance between wave packets in the quantum phase space, and black lines are the distance between the expectation values of operators $\hat q,\hat p$. (a)  $K=0.3$, $m=30$; (b) $K=0.9$, $m=30$; (c) $K=1.5$, $m=30$.   (d) has the same parameters  $K=1.5$, $m=30$ but with a longer time evolution.}
\end{figure}

\begin{figure}[htbp]
  \includegraphics[width=\columnwidth]{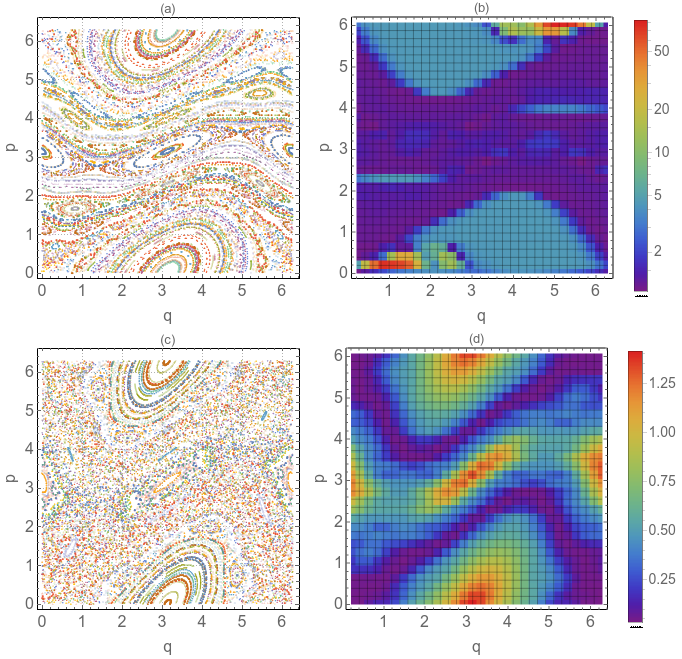}
  \caption{\label{fig2}(a)(c) Classical Poincar\'e sections  and (b)(d) quantum chaos measures  in the quantum phase space. The chaos measure for a Planck cell
  is computed by the evolution of a quantum state which is initially localized at the Planck cell.
  The figures are plotted by scanning the entire quantum phase space. Note that plotted in  (b) is colored with logarithm scaling and the original values range is $[1.07, 82.3]$. In (d), the values of chaos measure  range in $[0.03, 1.41]$.}
\end{figure}

In our numerical calculation, we choose the initial quantum states localized at $(4.7,3)$ and $(4.7+2\pi/m,3+2\pi/m)$. The dynamics
near these two points becomes chaotic as $K$ increases as shown in  Fig.\ref{fig2}(a,c).
The initial quantum states are the maximally localized Gaussian wave packets of the following form
\begin{equation}
  \langle x\ket{\psi(x_0, p_0)} =\frac {1} {(2\pi \hbar_{\text{eff}})^{1/4}} \exp\Big[ - \frac {(x-x_0)^2} {2\hbar_{\rm eff}}  + \frac {\text{i} xp_0} {\hbar_{\rm eff}}\Big]
\end{equation}
As the two states evolve with time,  we compute numerically the physical distance between them and see how they change with time.
For comparison, we have also computed two other distances. One is the distance between the expectation values of $\hat q, \hat p$ for these two different quantum states. For convenience, we call it expectation distance. The other is the distance between two corresponding classical trajectories starting at $(4.7,3)$ and $(4.7+2\pi/m,3+2\pi/m)$. The results are plotted in Fig.\ref{fig1}.

It is clear from Fig.\ref{fig1} that the physical distance agrees very well with the classical distance for the first several kicks. In contrast,  during these kicks,
the expectation distance can deviate largely from the classical distance. As the evolution goes on, both physical distance and expectation distance
deviate far from the classical distance as expected because the wave packets get distorted.
When $m$ is sufficiently large or, equivalently, $\hbar_{\rm eff}$ is small enough, the Ehrenfest time would be longer than the Lyapunov time.
In these cases, we should see that the physical distance agree with the classical distance for a much longer time; as a result, we  would
be able to estimate numerically the quantum Lyapunov exponent. Unfortunately, due to our limited computation power,
we can not compute for very large $m$.

We have also computed quantum chaos measure for the kicked rotor.
We use the maximally localized Gaussian wave packets as the initial states, scan the entire quantum phase space, and compute  the measure
for each Planck cell.  The results are plotted in Figs. \ref{fig2}(b,d) and compared to the classical Poincar\'e sections in Figs. \ref{fig2}(a,c).
The resemblance between them is unmistakeable.
Note that the chaos measure is the distance between long-time averaged density matrix and the maximally mixed state.
In Figs. \ref{fig2}(b), the values of our measure are very large for many Planck cells. This means that the wave packets starting at these Planck cells
do not spread out much and stay near the original Planck cells (e.g. see the upper panels in Fig. \ref{kicked}). 
When the kicking strength $K$ is large, the values of the measure become much smaller,
indicating that the quantum dynamics become more chaotic and the wave packets began to spread out to 
the large portions of the phase space (e.g. see the lower panels in Fig. \ref{kicked}).


\subsection{\label{subsec:5-2}Three-site Bose-Hubbard Model}
We consider a different system,  a three-site Bose-Hubbard model described by the following Hamiltonian~\cite{Han2016}
\begin{equation}
  \hat H = -\frac {c_0}{2} \sum_{\substack{1\leq i,j\leq 3 \\ i\neq j}} \hat a_i^\dagger \hat a_j + \frac c {2N}
  \sum_{j=1}^3 \hat a_j^\dagger \hat a_j^\dagger \hat a_j \hat a_j \,,
  \label{bh-H}
\end{equation}
where $\hat{a}_j^\dagger$ and $\hat{a}_j$ are the Bosonic creation and annihilation operators for the mode $j$. $c$ is the scaled interaction strength
and $N$ is the number of Bosons in the system.
This system has a mean-field limit at $N\rightarrow \infty$, whose  Hamiltonian is
\begin{equation}
  H_{\text{mf}} = -\frac {c_0}{2} \sum_{\substack{1\leq i,j\leq 3 \\ i\neq j}} a_i^* a_j + \frac c 2 \sum_{j=1}^3 |a_j|^4\,, \label{bh-H-cl}
\end{equation}
where $|a_1|^2+|a_2|^2+|a_3|^2=1$. In this Bose-Hubbard model, the quantumness is controlled by the particle number $N$ and
 the mean field Hamiltonian is its ``classical counterpart".  As we will show,  the physical distance
 is still applicable in this type of systems.

Since in the mean-field model each mode $a_j$ has a definite amplitude and phase, we choose a basis $\B$ for the quantum model where each basis vector
contains information for both amplitude (or particle number) and phase. For convenience,  we take the total particle number $N=L^2-1$, where $L$ is an integer.
The basis vector in $\B$ is denoted as
$\ket{\ell_1,\vartheta_1;\ell_2,\vartheta_2}$; its expectations for particles numbers are $\ell_{1,2}L+(L-1)/2$ and for phases $2\pi \vartheta_{1,2}/L$.
The details of this  orthonormal  basis
$\{\ket{\ell_1,\vartheta_1;\ell_2,\vartheta_2}\}$ with $0\le \ell_{1,2}, \vartheta_{1,2}\le L-1$ can be found in Appendix \ref{app-1}.
This effectively creates  a 4-dimensional quantum phase space with $L\times L\times L\times L$ Planck cells.
Therefore, a natural choice for the basis metric is
\begin{equation}
  d(\{\ell,\vartheta\},\{\ell',\vartheta'\}) = \frac 1 L \sqrt{\sum_{i=1}^2 \big(\Delta \ell_i^2 + \Delta \vartheta_i^2\big)}
\end{equation}
where $\Delta\ell_i = |\ell_i - \ell'_i|$ and $\Delta\vartheta_i = \min\{ |\vartheta_i - \vartheta'_i|, L - |\vartheta_i - \vartheta'_i|\}$($\vartheta_i$  is periodic).
We can compute the physical distance between two quantum states or two classical points according to this metric just like what we did for the kicked rotor.

The classical motion of the system is nonintegrable for generic $c$. This is evident in the Poincar\'e section for $c/c_0=2$, $E=0.8c_0$, $n_2=0.2475$, $\dot{n}_2>0$
shown in Fig.\ref{bh-cl-ps}, where we see both regular and chaotic motions. Note that $n_{1,2}=|a_{1,2}|^2$ and $\theta_{1,2}={\rm arg} a_{1,2} - {\rm arg} a_{3}$.
We choose the quantum initial state to be a coherent state $| \Psi \rangle=\frac{1}{\sqrt{N!}}\left(\sum_{i=1}^{3}a_i\hat{a_i}^{\dagger}\right)^N |0\rangle$;
its shape in our quantum phase space will be close to a Gaussian packet centered at the classical point $ (a_1, a_2, a_3)^{\mathrm{T}} $
with a width of order $\sqrt{1/N}\sim 1/L$ \cite{Han2016}.

\begin{figure}[htbp]
\centering
\subfloat{\label{bh-Lya-R}
\includegraphics[width=0.95\columnwidth]{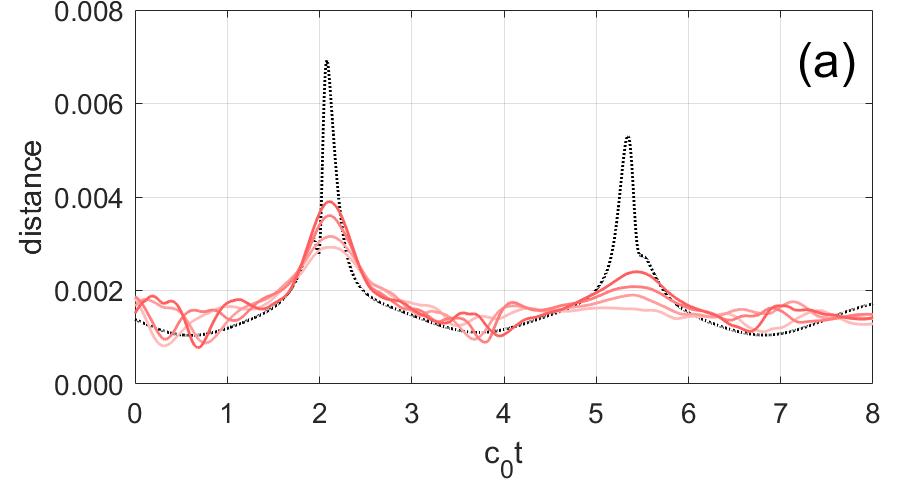}
}\\
\subfloat{\label{bh-Lya-C}
\includegraphics[width=0.95\columnwidth]{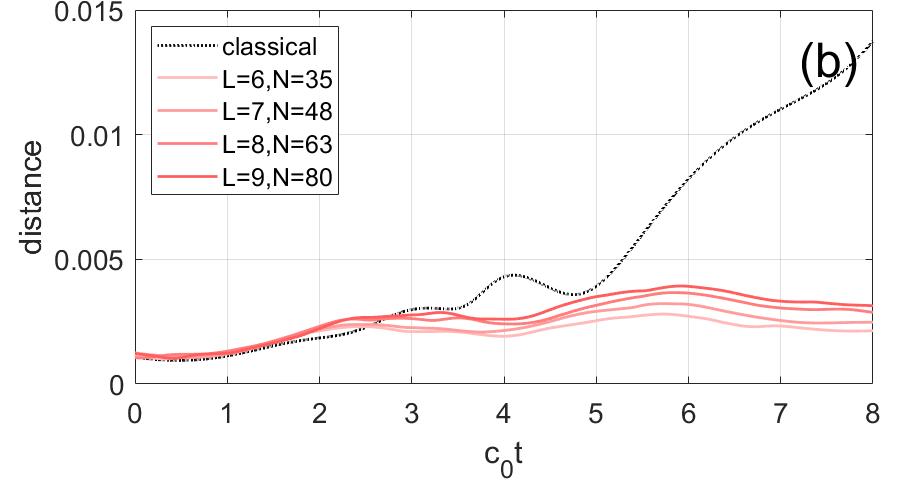}
}
\caption{Physical distances between two pairs of perturbed initial conditions: (a) regular case, $(n_1, \theta_1)=(0.220, 0.8\pi)$ and $(0.221, 0.8\pi)$;
(b) chaotic case, $(n_1, \theta_1)=(0.420, 0.8\pi)$ and $(0.421, 0.8\pi)$. Both are chosen from the Poincar\'e section in Fig.\ref{bh-cl-ps}.
The regular classical distance is well recovered by quantum distances for time up to $c_0t=8$, and the sharp peaks in the classical distance curve
is also prominent in quantum curves. As the quantum phase space resolution $L$ is increased, the peaks become closer
and closer to the classical one. In the chaotic case, although the quantum distance start deviating from the classical
distance at an earlier stage due to a fast Ehrenfest breakdown, its smoothness stands in sharp contrast to
the rich structure in regular classical distance curves as well as regular quantum distance curves,
demonstrating the stark difference  between regular motion and chaotic motion. }
\label{bh-Le}
\end{figure}

\begin{figure}[htbp]
\centering
\subfloat{
\includegraphics[width=0.5\columnwidth]{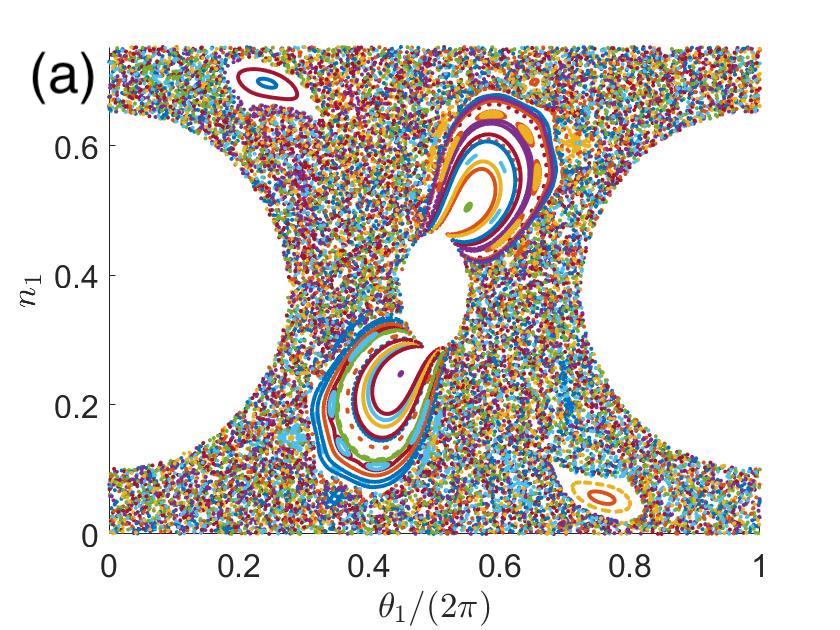}
\label{bh-cl-ps}
}
\subfloat{
\includegraphics[width=0.5\columnwidth]{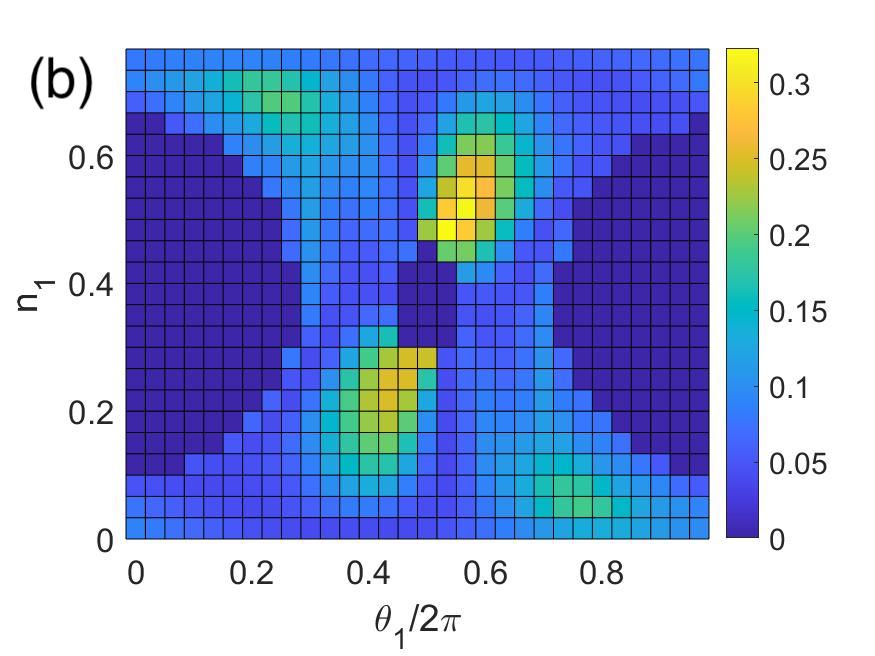}
\label{bh-C-I}
}
\caption{(a) Classical Poincar\'e section of the mean-field model (see Eq.(\ref{bh-H-cl})) at $c/c_0=2$, $E=0.8c_0$, $n_2=0.2475$, $\dot{n}_2>0$.
(b) The corresponding quantum chaos measure with $L=10$ (see main text for computation details).
Note that the side length of little squares in (b) is equal to 1/(3L), the sampling step length, instead of the inverse resolution $1/L$.  It is clear
that the yellow and green areas, where the quantum chaos measure is large,   coincide with the classical regular islands while
 the blue areas, the quantum chaos measure is small,  correspond to the chaotic sea. }
\label{bh-ci}
\end{figure}

We first choose a pair of initial conditions, $(n_1, \theta_1)=(0.220, 0.8\pi)$ and $(n_1, \theta_1)=(0.221, 0.8\pi)$, which are
located in the integrable island of  the Poincar\'e section in Fig.\ref{bh-cl-ps}.
The results are presented in Fig.\ref{bh-Lya-R}.
It can be seen that, when the dynamics is regular, the quantum physical distance coincides with the classical distance very well in a relatively long period of time.
When  the quantum resolution $L$ is increased from 6 to 9, the quantum physical distances have an obvious inclination to converge to the classical distance.
This is quite surprising because the difference between two initial conditons is only of order $10^{-3}$ while for the highest quantum resolution $L=9$
in our simulation  the size of the Planck cell is of order $10^{-1}$. So, we expect that  our physical distance match classical distance  well even
when the size of the Planck cell is significantly smaller than the classical distance between two initial conditions.

We choose a different pair of  initial conditions, $(n_1, \theta_1)=(0.420, 0.8\pi)$ and $(n_1, \theta_1)=(0.421, 0.8\pi)$,
which are located in the chaotic sea of Fig.\ref{bh-cl-ps}. The numerical results are shown in Fig. \ref{bh-Lya-C}.
In this chaotic case, the total time that the quantum physical distances and the classical distance coincide is much shorter.
This can be explained with the Ehrenfest time, the time scale when the quantum-classical dynamics breakdown.
For the chaotic dynamics, the Ehrenfest time is short and proportional to $\ln N$\cite{Han2016}, while for the integrable dynamics
this timescale is much longer and proportional to $\sqrt{N}$ \cite{Zhao2019}. So, for this chaotic dynamics, to see numerically exponential divergence of the quantum physical distance,
we have to have $N$ (or $L$) exponentially large. Unfortunately, for both integrable and chaotic cases, large $N$ is beyond our numerical capacity.

We also computed quantum chaos measure for this Bose-Hubbard system, corresponding to the classical Poincar\'e section in Fig.\ref{bh-cl-ps} with $c/c_0=2$.
We divide the phase space with the quantum resolution $L=10$, that is, the total particle number being $N=99$.
Our initial quantum states are coherent states $| \Psi_c(t=0) \rangle$, which are localized wavepackets occupying $O(1)$ Planck cells. We calculate their long-time average density matrix $\rho_{\infty} =\lim_{T\to\infty} \frac{1}{T}\int_0^T \mathrm{d} t |\Psi_c(t)\rangle \langle \Psi_c(t) |$, and project them onto the quantum phase space and obtain the distribution $P_c(\ell_i,\vartheta_i)=\mathrm{Tr} \{ \rho_{\infty} |\ell_i,\vartheta_i \rangle \langle \ell_i,\vartheta_i | \}$. The classical (or mean-field) dynamics is  limited to
a constant energy surface in the phase space. In contrast, the quantum dynamics is limited to an energy shell with certain thickness.
To effectively reduce the computation burden, we only pick out Planck cells located within the Gaussian-broadening energy shell. To be specific, we Gaussian fit the smoothened envelope of the energy spectrums of these selected coherent states, with the fitting goodness $R^2=0.994$, and select out the phase cells with energy expectation values
within $\pm 3\sigma$ of the Gaussian. This method is able to capture $70\%\pm 7\%$ of the original coherent packets. We then set the energy shell with
this Gaussian envelope to be the ergodic reference $\rho_{\rm erg}$ of the system following \cite{Heller1987}, in place of the classical microcanonical ensemble, and project it onto the space $P_{\rm erg}(\ell_i,\vartheta_i)=\mathrm{Tr} \{ \rho_{\rm erg} |\ell_i,\vartheta_i \rangle \langle \ell_i,\vartheta_i | \}$. Then we calculate the physical distance between $P_{\rm erg}(\ell_i,\vartheta_i)$ and each $P_c(\ell_i,\vartheta_i)$ and obtain the chaos measure. The results are plotted in Fig.\ref{bh-C-I}.
We see that the classical Poincar\'e section in Fig.\ref{bh-cl-ps} is very well recovered. The regular islands are distinguished by the coherent initial states that have a large physical distance from the ergodic envelope, while the chaotic sea is filled with initial states that are very close in physical distance to the ergodic envelope. Quite surprisingly, even the two small regular islands with size of only one single quantum phase cell are clearly seen. Therefore, we expect our chaos measure proposed here is able to distinguish regular island structures with size no smaller than the order of one single Planck cell.

\subsection{\label{subsec:5-3}Spin Chain}
\begin{figure}
  \includegraphics[width = 0.9\columnwidth]{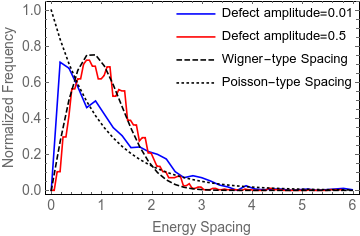}
  \caption{Eigen-energy spacing statistics for the XXZ spin chain that has $15$ spins in the sub-Hilbert space with $5$ spins up.
  The defection is on the site $i^* = 2$ and the boundary condition is open with $J_1 = 1.0$,  $J_2 = 0.5$.
  The red line  is for $\epsilon = 0.5$ and is close to the Wigner-Dyson distribution (dashed line) while the blue line is for
   $\epsilon = 0.01$ and is close to Poison distribution(dotted line).}
  \label{fig5}
\end{figure}

We now study a system which does not have a clear classical counterpart. It is the spin-1/2 XXZ model
with disorder described by the following Hamiltonian~\cite{Gubin2012}
\begin{equation}
  \hat H= \sum_{i=0}^{N-1} h_i \hat s_i^z + \sum_{i=0}^{N-2}\Big\{ J_1\big(\hat s_i^x \hat s_{i+1}^x + \hat S_i^y \hat s_{i+1}^y\big) +
  J_2 \hat s_i^z \hat s_{i+1}^z\Big\}\,,
\end{equation}
where $\hat s_i^{x,y,z}$ are spin operators at the $i$th site and $h_i = \epsilon \delta_{i,i^*}$ is the magnetic field at a given random site denoted by $i^*$.
In our model, $S^z=\sum_{i=1}^N s^z_i$ is conserved and the Hilbert space can be divided into subspaces labelled by $S^z$.
The spin system has different eigen-energy spacing statistics with different values of $\epsilon$~\cite{Gubin2012}. Two examples are
shown in Fig. \ref{fig5}, which show that  the case $\epsilon=0.01$ is largely integrable while the case $\epsilon=0.5$ is chaotic.

To compute the quantum chaos measure, we choose the set of common eigenstates of all $\hat s_i^z$ as the basis $\B$ and denote them as $0,1$ valued vectors $\{\ket{s_1,s_2,\cdots,s_N}\}_{s_i\in \{0,1\}}\}$.  The distance between them is defined as the $L_1$ measure between
the arrays of positions of $1$s. For example, the array for the positions of $1$s in the state $\ket{1,1,0,0,0}$ is $(0,1)$, and the array
for $\ket{0,0,0,1,1}$ is $(3,4)$. So,  the distance between them is
\begin{equation}
  d(\ket{1,1,0,0,0},\ket{0,0,0,1,1}) = |3-0| + |4-1| = 6
\end{equation}
Note that this metric is different from the Hamming distance, which is $4$ between the states $\ket{1,1,0,0,0}$ and $\ket{0,0,0,1,1}$.
In fact,  this metric is the same as our distance defined for the  many-body states in Sec. \ref{sec:3}
if we treat the states $\ket{s_1,\cdots,s_N}$ as Fock states for Fermions with $s_j$ particle in the single particle mode $\ket{j}$ and define
the distance between corresponding single particle states $\ket{j}$'s as $d(\ket{i},\ket{j}) = |i-j| \ ; \ i,j=1,\cdots,N$. For example,
the most efficient way to transport $(1,1,0,0,0)$ to $(0,0,0,1,1)$ is to move each \(1\) in the first array to the position of corresponding \(1\)s one by one in the second array.

\begin{figure}[htbp]
  \includegraphics[width = 0.9\columnwidth]{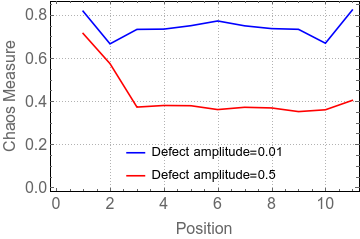}
  \caption{Quantum chaos measures for the $10$ initially localized states in the XXZ spin chain model.
  The green line is for  $\epsilon=0.01$ and the red line for $\epsilon=0.5$.
  The left side of the red line is large is caused by the defection on the second site $i^*=2$. The overall decrease of the chaos measure
  from $\epsilon=0.01$ to $\epsilon=0.5$ shows the spin chain becomes more
  chaotic as the defection increases, which agrees the results of energy spacing statistics. }
  \label{fig6}
\end{figure}

In our numerical computation, we choose $N=15$ and focus on the subspace with  $5$ spins up.
 The initial localized states are chosen to be states that have consecutive $1$s in their boolean-valued arrays, such as $(1,1,1,1,1,0,\cdots,0)$ and $ (0,1,1,1,1,1,\cdots)$.
 There are in total  ten of  them, which are numbered according to the position of the first $1$. The computed quantum chaos measure is shown in Fig.\ref{fig6}.
 It is clear from the figure that the chaos measure is much smaller for the chaotic case $\epsilon=0.5$ than for the non-chaotic case $\epsilon=0.01$.
 This is in agreement with the eigen-energy spacing statistics shown in Fig. \ref{fig5}.

\section{\label{sec:6}Discussion and Conclusion}

In some cases the Wasserstein distance is not robust with respect to the distance matrix $d_{ij}$ defined for a pair of basis vector.
For example, consider two distribution: $\delta(x)$ and $(1-\eta) \delta(x) + \eta \delta(x-d)$ on a one dimensional Euclidean space.
The Wasserstein distance between them is nothing but $\eta d$. That means no matter how small $\eta$ is, one can
find a sufficient large $d$ so that the Wasserstein distance diverges. That is what we meant by
that  the distance is not robust with respect to $d_{ij}$ . Fortunately, the distance matrix $d_{ij}$
in our numerical examples have upper bounds  so that we do not need to worry about  this.

Another challenge for our proposed physical distance is the complexity of computing the Wasserstein distance.
In our numerical implement, we use python module named pyemd to compute the Wasserstein distance \cite{pele2008,pele2009}.
Though the convex optimization problem for computing the distance is easy (it is a linear programing problem in discretized form\cite{oberman2015efficient}), we are facing an
exponentially high dimensional Hilbert space in quantum mechanics and our code can not handle larger systems.
But the core of this challenge is the dimension of the Hilbert space, which should  also be a challenge to  any other definitions with the similar concept.
In this sense, our definition has an equivalent complexity to others.

In conclusion, we have shown that there is genuine quantum chaos despite that quantum dynamics is linear. 
This is revealed with the physical distance that we 
proposed for two quantum states. This quantum distance is based on the Wasserstein distance between two probability distributions. We call it "physical" because it faithfully measures the difference of physical properties.
This physical distance can  be very small for two orthogonal initial quantum states, and then diverge exponentially 
during the ensuing quantum  chaotic motion. With physical distance, we have defined two parameters to 
characterize the quantum dynamics: quantum Lyapunov exponents for the short time dynamics and quantum chaos measure
for the long time dynamics. The latter allows us
to construct the quantum analogue of the classical Poincar\'e section, where regions for 
regular quantum motions and chaotic quantum motions are mapped out.

\acknowledgements
This work is supported by the the National Key R\&D Program of China (Grants No.~2017YFA0303302, No.~2018YFA0305602),
National Natural Science Foundation of China (Grant No. 11921005), and
Shanghai Municipal Science and Technology Major Project (Grant No.2019SHZDZX01).


\clearpage

\appendix
\section{\label{app-1}Quantum phase space for the three-site Bose-Hubbard model}
\subsection{Construction of the quantum phase space}
The total particle number $N$ is conserved in the three-site Bose-Hubbard Hamiltonian (Eq. \ref{bh-H}). As a result,
in  the Fock states $|N_1,N_2,N_3\rangle$, there are only two free parameters $N_{1,2}$ and $N_3=N-N_1-N_2$.
We consider a Hilbert space spanned by  Fock states $|N_1,N_2\rangle$ with $0\le N_{1,2}\le N$, which contains
the Hilbert space of the three-site Bose-Hubbard system. We further take $N=L^2-1$. For this $L^4$-dimensional Hilbert space,  we define a set of orthonormal basis
\begin{eqnarray}\label{bh-Wb-def}
&& |\ell_1,\vartheta_1;\ell_2,\vartheta_2\rangle =\\
&&\frac{1}{L}\sum_{N_1=0}^{L-1}\sum_{N_2=0}^{L-1} e^{i\frac{2\pi}{L}(N_1\vartheta_1+N_2\vartheta_2)} |N_1+\ell_1 L, N_2+\ell_2 L\rangle\,,\nonumber
\end{eqnarray}
where the four indices $\ell_{1,2}, \vartheta_{1,2} = 0,1,..., L-1$.  As a result, the $L^4$-dimensional Hilbert space is arranged into
a 4-dimensional phase space which is divided into  $L\times L\times L\times L$ Planck cells. And each Planck cell is
represented by $\ket{\ell_1,\vartheta_1;\ell_2,\vartheta_2}$. For this 4-dimensional phase space, there are two pairs of conjugate observables,
$\hat{N}_{1,2}$, the particle number operators , and $\hat{\theta}_{1,2}$, the relative phase  operators. They can be defined as
\begin{eqnarray}
\hat{N}_i &=&\sum_{N_1=0}^N\sum_{N_2=0}^N N_i|N_1,N_2\rangle \langle N_1,N_2|, \\
\hat{\theta}_i&=&\sum_{M_1=0}^N\sum_{M_2=0}^N \theta_{M_i} | \theta_{M_1},\theta_{M_2} \rangle \langle \theta_{M_1},\theta_{M_2}|,
\end{eqnarray}
where $| \theta_{M_1},\theta_{M_2}\rangle$ is the Fourier transformation of the Fock basis
\begin{equation}
| \theta_{M_1},\theta_{M_2}\rangle=\frac{1}{N+1}\sum_{N_1=0}^{N} \sum_{N_2=0}^{N}e^{i(N_1\theta_{M_1}+N_2\theta_{M_2})}|N_1,N_2\rangle
\label{bh-ntheta}
\end{equation}
with $\theta_{M_i}=\theta_i^{(0)}+2\pi M_i/(N+1)$ and
$M_i=0,1,...,N$. In light of this, we can recast Eq.(\ref{bh-Wb-def}) into
\begin{eqnarray}\label{bh-Wb-def2}
&& |\ell_1,\vartheta_1;\ell_2,\vartheta_2\rangle =\\
&&\frac{1}{L^3}\sum_{M_1=0}^{N}\sum_{M_2=0}^{N} \frac{1-e^{-iL(\theta_{M_1}-\frac{2\pi\vartheta_1}{L})}}{1-e^{-i(\theta_{M_1}-\frac{2\pi\vartheta_1}{L})}} \frac{1-e^{-iL(\theta_{M_2}-\frac{2\pi\vartheta_2}{L})}}{1-e^{-i(\theta_{M_2}-\frac{2\pi\vartheta_2}{L})}}  \nonumber \\
&&~~~~~~~~~~~~~\times e^{-iL(\ell_1 \theta_{M_1} + \ell_2 \theta_{M_2})} |\theta_{M_i}, \theta_{M_2}\rangle   \nonumber
\end{eqnarray}
where each fraction takes its limit value if its denominator is 0. \par
We will show in the follow that each $|\ell_1,\vartheta_1;\ell_2,\vartheta_2\rangle$ state represents a Planck cell in the phase space in the sense that their positions are fixed by the four parameters $\ell_i,\vartheta_i$, and that their shapes are localized.

\subsection{The positions of the Planck cells}
We will analytically verify that for
a given Planck cell $|\ell_1,\vartheta_1;\ell_2,\vartheta_2\rangle$, $\ell_{i}$ is proportional to the  expectation value of particle number at this cell
and $\vartheta_{i}$ is proportional to the  expectation value of the phase, up to some correction terms. The expectations are
\begin{eqnarray}
\langle\hat{N}_i\rangle_{\ell_i,\vartheta_i}&=&\ell_{i} L + \frac{L-1}{2},\label{bh-N-mean} \\
\langle\hat{\theta}_i\rangle_{\ell_i,\vartheta_i}&=&\frac{2\pi}{L}\vartheta_{i} + \frac{1}{L^3}\sum_{M_i=0}^N \left|\frac{\sin \frac{L\tilde{\theta}_{M_i}}{2}}{\sin\frac{\tilde{\theta}_{M_i}}{2} }\right|^2 \tilde{\theta}_{M_i}, \label{bh-theta-mean}
\end{eqnarray}
where $\langle \cdot \rangle_{\ell_i,\vartheta_i}$ denotes the expectation value of the state $|\ell_1,\vartheta_1;\ell_2,\vartheta_2\rangle$, and $\tilde{\theta}_{M_i}=\theta_{M_i}-\frac{2\pi}{L}\vartheta_i$. The normalization of Eq.(\ref{bh-Wb-def2}) has been used in deriving Eq.(\ref{bh-theta-mean}). \par

\begin{figure}[htbp]
\centering
\subfloat{\label{fig:mean_theta_1}
\includegraphics[width=0.48\columnwidth]{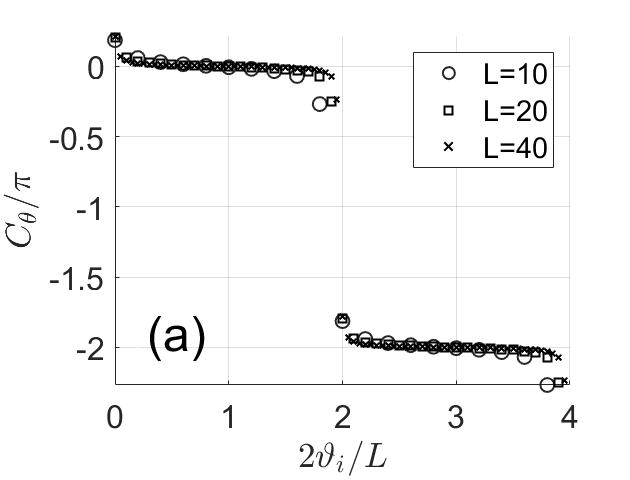}
}
\subfloat{\label{fig:mean_theta_s}
\includegraphics[width=0.48\columnwidth]{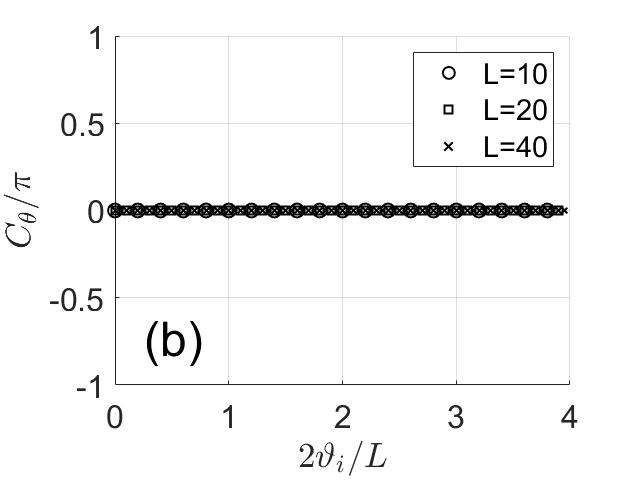}
}
\caption{\label{fig:mean_theta} The values of the correction term in Eq.(\ref{bh-theta-mean}), $C_{\theta}=\frac{1}{L^3}\sum_{M_i=0}^N \left|\frac{\sin \frac{L\tilde{\theta}_{M_i}}{2}}{\sin\frac{\tilde{\theta}_{M_i}}{2} }\right|^2 \tilde{\theta}_{M_i} $.
(a) $\theta_i^{(0)}$ is fixed at $-\pi/L$. (b) $\theta_i^{(0)}$ is changed  to keep all the $\tilde{\theta}_{M_i}$ lie in the region of $[-\pi, \pi)$.
Two periods of $\vartheta_i$ are plotted in these two figures. }
\end{figure}

We can further show that the correction terms can be regarded as constants independent of $\vartheta_i$. In Eq.(\ref{bh-N-mean}), this is obvious. We only need to
examine Eq.(\ref{bh-theta-mean}). We notice that in Eq.(\ref{bh-theta-mean}), with fixed $\theta^{(0)}_i$, altering $\vartheta_i \to \vartheta_i+1$ is equivalent to adding an additional term to $\theta_{M_i}\to \theta_{M_i} -2\pi/L$, and keeping $\vartheta_i$ unchanged,
\begin{eqnarray}
 &&\frac{1}{L^3}\sum_{M_i=0}^N \left|\frac{\sin \frac{L\tilde{\theta}_{M_i}}{2}}{\sin\frac{\tilde{\theta}_{M_i}}{2} }\right|^2 \tilde{\theta}_{M_i}   \\
 && \to \frac{1}{L^3}\sum_{M_i=0}^N \left|\frac{\sin \frac{L(\tilde{\theta}_{M_i}-2\pi/L)}{2}}{\sin\frac{\tilde{\theta}_{M_i}-2\pi/L}{2} }\right|^2 (\tilde{\theta}_{M_i} -2\pi/L) \nonumber \\
 && =\frac{1}{L^3}\sum_{M_i=0}^{N} \left|\frac{\sin \frac{L\tilde{\theta}_{M_i-L}}{2}}{\sin\frac{\tilde{\theta}_{M_i-L}}{2} }\right|^2 \tilde{\theta}_{M_i-L}\nonumber \\
 && = \frac{1}{L^3}(\sum_{M_i=0}^{N-L} + \sum_{M_i=-L}^{-1}) \left|\frac{\sin \frac{L\tilde{\theta}_{M_i}}{2}}{\sin\frac{\tilde{\theta}_{M_i}}{2} }\right|^2 \tilde{\theta}_{M_i} \nonumber\\
 &&= \frac{1}{L^3}\sum_{M_i=0}^{N-L} \left|\frac{\sin \frac{L\tilde{\theta}_{M_i}}{2}}{\sin\frac{\tilde{\theta}_{M_i}}{2} }\right|^2 \tilde{\theta}_{M_i} \nonumber\\
 &&~~ +\frac{1}{L^3} \sum_{M_i=N-L+1}^{N} \left|\frac{\sin \frac{L(\tilde{\theta}_{M_i}-2\pi)}{2}}{\sin\frac{\tilde{\theta}_{M_i}-2\pi}{2} }\right|^2( \tilde{\theta}_{M_i} -2\pi)\nonumber\\
 && = \frac{1}{L^3}\sum_{M_i=0}^{N} \left|\frac{\sin \frac{L\tilde{\theta}_{M_i}}{2}}{\sin\frac{\tilde{\theta}_{M_i}}{2} }\right|^2 \tilde{\theta}_{M_i}  - \sum_{M_i=N-L+1}^{N} \frac{2\pi}{L^3}\left|\frac{\sin \frac{L\tilde{\theta}_{M_i}}{2}}{\sin\frac{\tilde{\theta}_{M_i}}{2} }\right|^2 \nonumber
\end{eqnarray}
We can see that there are special  points  $\tilde{\theta}_{M_i}\equiv 0 ({\rm mod} 2\pi)$ where $\left|\frac{\sin \frac{L\tilde{\theta}_{M_i}}{2}}{\sin\frac{\tilde{\theta}_{M_i}}{2} }\right|^2$ is of order $O(L^2)$. Otherwise,  the alteration of the correction term induced by $\vartheta_i \to \vartheta_i+1$
is of order $O(L\cdot L^{-3})=O(L^{-2})$, which is ignorable compared to the shift in the first term of Eq.(\ref{bh-theta-mean}), ${2\pi}/L$.
Since $\vartheta_i$ has a period of $L$ and $\sin\tilde{\theta}_{M_i}/2=\sin(\theta_{M_i}-2\pi\vartheta_i/L)/2 $ approaches 0 for only once over the entire period,
we can always avoid those points in an entire period by carefully choosing $\theta_i^{(0)}$. Therefore, we can say that the correction term in Eq.(\ref{bh-theta-mean})
is approximately a constant. This point is illustrated in Fig.\ref{fig:mean_theta_1}.  \par
\begin{figure}[htbp]
\centering
\subfloat{\label{fig:amp1}
\includegraphics[width=0.48\columnwidth]{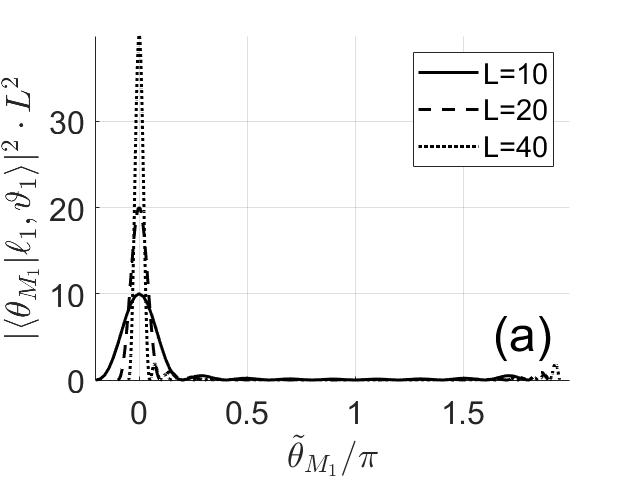}
}
\subfloat{\label{fig:amp2}
\includegraphics[width=0.48\columnwidth]{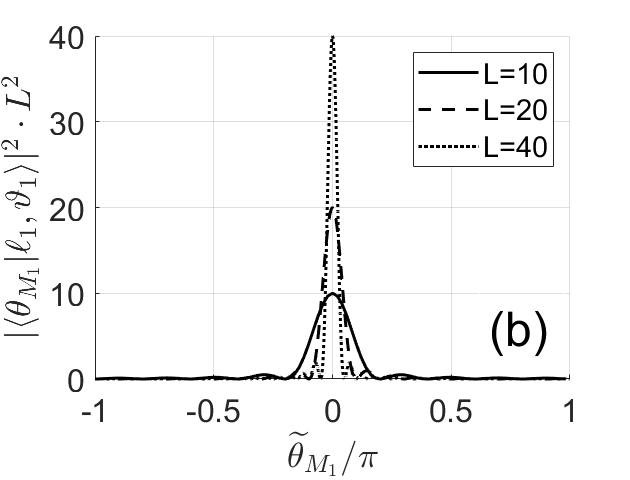}
}
\caption{\label{fig:amp}  $|\langle \theta_{M_1} | \ell_1,\vartheta_1 \rangle|^2$ versus $\tilde{\theta}_{M_1}$, with fixed $\vartheta_1=0$ and different choice of $\theta_1^{(0)}$. (a) $\theta_1^{(0)}=-\pi/L$. (b) $\theta_1^{(0)}=-\pi$. }
\end{figure}

We can understand this question from another point of view. Since expectation values of well 1 and 2 are independent of each other, we may ignore the degrees of freedom related to well 2 and consider a reduced double-well case. Note that $\theta_1$ is actually a periodic quantity, and we should actually plot the amplitude $|\langle \theta_{M_1} | \ell_1,\vartheta_1 \rangle|^2$ on a ring. Therefore, different choice of $\theta_1^{(0)}$ turns out to represent cutting the ring at different positions,
as illustrated by Fig.\ref{fig:amp}. Obviously Fig.\ref{fig:amp1} will have a slightly larger expectation value of $\hat{\theta_1}$ than Fig.\ref{fig:amp2}, but as long as the cut is not in the peak, whose width scales as $O(L^{-1})$, the deviation will be small. This is the origin of the correction term in Eq.(\ref{bh-theta-mean}). \par
If we allow $\theta_1^{(0)}$ to change with the shifting of $\vartheta_1$ and keep all the $\tilde{\theta}_{M_1}=\theta_1^{(0)}+2\pi M_1/L^2-2\pi\vartheta_1/L$ ($M_1=0,1,...,L^2-1$) lie in the region of $[-\pi, \pi)$, as in Fig.\ref{fig:amp2}, then the peak in $|\langle \theta_{M_1} | \ell_1,\vartheta_1\rangle|^2$ will always be at the middle point of the region, hence the correction term will be exactly 0 for arbitrary $\vartheta_1$, as shown in Fig.\ref{fig:mean_theta_s}. \par

\subsection{Localization of the Planck cells}
Finally, we  analyze the fluctuations of these expectation values, i.e. the localization of the shapes of these Planck cells. The fluctuations are
\begin{eqnarray}
\Delta n_i&=&\frac{\sqrt{\langle\hat{N_i}^2\rangle_{\ell_i,\vartheta_i} -\langle\hat{N_i}\rangle_{\ell_i,\vartheta_i}^2 }}{N}\nonumber \\
&&=\frac{1}{\sqrt{12N}}\sim \frac{1}{\sqrt{12}L}, \\
\Delta \theta_i &=& \sqrt{\langle\hat{\theta_i}^2\rangle_{\ell_i,\vartheta_i} -\langle\hat{\theta_i}\rangle_{\ell_i,\vartheta_i}^2} \nonumber \\
&=&\sqrt{ \frac{1}{L^3}\sum_{M_i=0}^N \left|\frac{\sin \frac{L\tilde{\theta}_{M_i}}{2}}{\sin\frac{\tilde{\theta}_{M_i}}{2} }\right|^2 \tilde{\theta}_{M_i}^2 - C_{\theta}^2} \,.\label{eq:fluc_theta}
\label{bh-Wb-var}
\end{eqnarray}
As discussed above, we can choose  $C_{\theta}=0$ so that Eq.\ref{eq:fluc_theta} can be further estimated as
\begin{eqnarray}\label{eq:fluc_theta_2}
\Delta \theta_i &\sim& \sqrt{ \frac{1}{L^3}\int_{-\pi}^{\pi} \mathrm{d}x \cdot L^2\cdot \left|\frac{\sin \frac{Lx}{2}}{\sin\frac{x}{2} }\right|^2 x^2 } \\
&=& \sqrt{ \frac{1}{L^3}\int_{-\pi}^{\pi} \mathrm{d}x \cdot L^2\cdot \sin^2 \frac{Lx}{2} \frac{ x^2}{\sin^2 \frac{x}{2}} } \nonumber \\
&\sim& \sqrt{ \frac{A}{L}\int_{-\pi}^{\pi} \mathrm{d}x \cdot \sin^2 \frac{Lx}{2}  }\nonumber\\
&\sim&\sqrt{ \frac{\pi A}{L}}\nonumber
\end{eqnarray}
where $A$ is a constant of order $O(1)$ introduced due to the fact that $x/\sin(x/2) \sim O(1)$ over the entire region of $[-\pi,\pi)$.
Therefore, for both the particle number and the phase, the fluctuations of the expectation values converge to 0 as $L$ goes to infinity. This establishes the localized `cell' picture of each Planck cell. \par

\section{\label{app-2}Examples of  computing the Wasserstein distance}
We use two examples to compute the Wasserstein distance between distributions. We consider two distributions
\begin{equation}
  p_A(x)=\begin{cases}1/5 & x< 5 \\ 0 & x\geq 5\end{cases}
\end{equation}
and
\begin{equation}
\ p_B(x)=\begin{cases}1/5 & x \text{ is even} \\ 0 & \text{otherwise}\end{cases}
\end{equation}
on the set $\B=\{0,1,\cdots,9\}$. We  want to compute the Wasserstein distance between them and the uniform distribution.
We choose the metric  on $\B$ as $d(x,y) = |x-y| \ ; \ x,y\in \B$. The Wasserstein distance can be computed as follows.

\begin{figure}[tbp]
  \includegraphics[width=0.85\columnwidth]{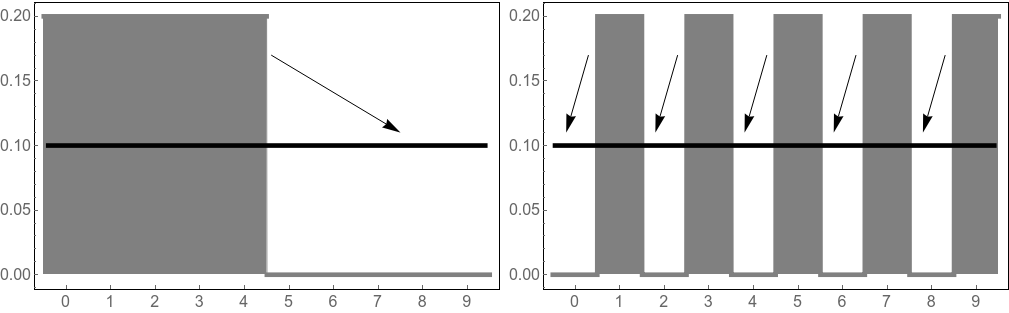}
  \caption{\label{fig-ap2}The most efficient way to achieve the transport between two distributions.}
\end{figure}
The transport matrix between $p_A$ and uniform distribution $p_E(x) = 1/10$ is $P_{ij} \in [0,1]$, which obeys
\begin{equation}
  \sum_{j\in \B} P_{ij} = p_A(i) \ ; \ \sum_{i\in \B}P_{ij} = p_E(j)
\end{equation}
The Wasserstein distance for \(\lambda=1\) is the minimum value of
\begin{equation}
  \sum_{i,j\in \B} P_{ij} d(i,j)\,.
\end{equation}
This shows that the Wasserstein distance is the most efficient way to transform one distribution to another.
As both distributions $p_A$ and $p_B$ are simple, the most efficient ways are  shown in Fig. \ref{fig-ap2}.
For the distribution $p_A$, the non zero  optimal \(P_{ij}^*\) are
\begin{equation}
  P_{i,i}^*=P_{i,i+5}^* = 1/10 \ ; \ i=0,1,\cdots,4
\end{equation}
which means the Wasserstein distance: $D_1(p_A,p_E)=5\times |5-0|\times 1/10 = 2.5$.
Similarly,  we have $D_1(p_B,p_E) = 5\times|1-0|\times 1/10=0.5$.

\end{document}